\documentclass[12pt,a4paper]{article}
\usepackage{graphics}
\usepackage{amsmath,amssymb}

\makeatletter
\newif\if@preliminary
\@preliminaryfalse
\def\preliminary{\@preliminarytrue}
\def\preprintno#1{\def\@preprintno{#1}}
\def\address#1{\def\@address{#1}}

\def\abstract#1{\def\@abstract{#1}}
\renewcommand\abstractname{ABSTRACT}
\newlength\preprintnoskip
\newlength\abstractwidth
\@titlepagetrue
\renewcommand\maketitle{\begin{titlepage}%
  \let\footnotesize\small
  \def\thefootnote{\fnsymbol{footnote}}
  \hfill\parbox{\preprintnoskip}{%
  \begin{flushright}\@preprintno\end{flushright}}\hspace*{1cm}
  \vskip 60\p@
  \begin{center}%
    {\Large\bf\boldmath \@title \par}\vskip 1cm%
    {\sc\@author \par}\vskip 3mm%
    {\@address \par}%
    \if@preliminary
      \vskip 2cm {\large\sf (PRELIMINARY DRAFT) \par \@date}%
    \fi
  \end{center}\par
  \@thanks
  \vfill
  \begin{center}%
    \parbox{\abstractwidth}{\centerline{\abstractname}%
    \vskip 3mm%
    \@abstract}
  \end{center}
  \end{titlepage}%
  \setcounter{footnote}{0}%
  \def\thefootnote{\arabic{footnote}}
  \let\thanks\relax\let\maketitle\relax
  \gdef\@thanks{}\gdef\@author{}\gdef\@address{}%
  \gdef\@title{}\gdef\@abstract{}\gdef\@preprintno{}
}%
\def\shortletter{%
 \setcounter{secnumdepth}{5}
 \def\paragraph{%
   \@startsection{paragraph}{4}{\parindent}%
     {3.25ex \@plus1ex \@minus.2ex}{-.5em}%
     {\reset@font\normalsize\bfseries}}%
 \renewcommand\theparagraph{\arabic{paragraph}.\hskip-.5em}
 \def\subparagraph{%
   \@startsection{subparagraph}{5}{\parindent}%
     {3.25ex \@plus1ex \@minus.2ex}{-.5em}%
     {\reset@font\normalsize\bfseries}}%
 \renewcommand\thesubparagraph{(\alph{subparagraph})\hskip-.5em}
}

\long\def\@makecaption#1#2{%
  \vskip\abovecaptionskip
  \sbox\@tempboxa{#1: \emph{#2}}%
  \ifdim \wd\@tempboxa >\hsize
    #1: \emph{#2}\par
  \else
    \hbox to\hsize{\hfil\box\@tempboxa\hfil}%
  \fi
  \vskip\belowcaptionskip}

\makeatother

\def\fmfL(#1,#2,#3)#4{\put(#1,#2){\makebox(0,0)[#3]{#4}}}

\newcounter{actr}

\newcommand{\fnnum}[1]{$^#1$}

\newcommand{\TeV}{\ensuremath{\textrm{TeV}}}
\newcommand{\GeV}{\ensuremath{\textrm{GeV}}}
\newcommand{\ab}{\ensuremath{\textrm{ab}}}
\newcommand{\fb}{\ensuremath{\textrm{fb}}}
\newcommand{\LL}{\mathcal{L}}
\newcommand{\tr}[1]{\textrm{tr}\left[#1\right]}

\begin{document}
\shortletter        
\preprintno{DESY--99--111\\TTP99--34\\MSU-HEP-90815\\hep-ph/9908409\\[.5\baselineskip]
 July 1999}
\title{%
 STRONGLY INTERACTING VECTOR BOSONS\\
 AT TeV $e^\pm e^-$ LINEAR COLLIDERS\\[.5\baselineskip]
 --- ADDENDUM ---
}
\vspace{0.3cm}
\author{%
 E.~Boos\fnnum1,
 H.--J.~He\fnnum2,
 W.~Kilian\fnnum3,
 A.~Pukhov\fnnum1,
 C.--P.~Yuan\fnnum2,
 and P.M.~Zerwas\fnnum4
}
\vspace{\baselineskip}
\address{
 \fnnum1{}Institute of Nuclear Physics, Moscow State University,\\
 119899 Moscow, Russia\\[.5\baselineskip]
 \fnnum2{}Department of Physics and Astronomy, Michigan State University,\\
 East Lansing, Michigan 48824, USA\\[.5\baselineskip]
 \fnnum3{}Institut f\"ur Theoretische Teilchenphysik, 
 Universit\"at Karlsruhe,\\
 D--76128 Karlsruhe, Germany \\[.5\baselineskip]
 \fnnum4{}Deutsches Elektronen-Synchrotron DESY,\\
 D--22603 Hamburg, Germany
}
\abstract{%
Extending earlier investigations, we analyze the quasi-elastic
scattering of strongly interacting electroweak bosons at 
high-energy $e^\pm e^-$
colliders.  The three processes $e^+e^-\to\bar\nu\nu W^+W^-$, $\bar\nu\nu
ZZ$ and $e^-e^-\to\nu\nu W^-W^-$ are examined at a c.m.\ energy of 
$1~\TeV$ for high-luminosity runs.  The expected experimental error on the
scattering amplitude, parameter-free to leading order in the chiral
expansion of the $WW$ interactions, is estimated for $1~\TeV$ colliders
at the level of ten percent, providing a stringent test of strong
interaction mechanisms for breaking the electroweak symmetries.
}
\maketitle

\baselineskip20pt   
\paragraph{}
Unitarity leads to the alternative scenarios that either a light Higgs
boson is realized in the electroweak sector of the Standard Model
(SM), or that the electroweak $W^\pm,Z$ gauge bosons become strongly
interacting at high energies~\cite{Uni}.  Within the canonical
formulation of the Standard Model, analyses of the high-precision
electroweak data are in striking agreement with the existence of a
light Higgs boson~\cite{data}.  However, if the SM interactions are
supplemented by low-energy remnants of new interactions at high energy
scales, alternatives to the light Higgs scenario are still viable
(see, e.g., Ref.\cite{BS99}).

In a preceding investigation~\cite{WW1} we have analyzed the
quasi-elastic scattering of $W^\pm,Z$ bosons,
\begin{equation}
  WW \to WW
\end{equation}
at \TeV\ $e^\pm e^-$ linear colliders in the high-energy range where
the strong interactions between the electroweak gauge bosons become
effective in the absence of a light Higgs boson.  The strong
interactions of the $W$~bosons can, in a natural way, be traced back
to the interactions of Goldstone bosons which are associated with the
spontaneous breaking of a chirally invariant theory, characterized by
an energy scale $\Lambda\sim\mathcal{O}(1~\TeV)$.  As formulated by
the equivalence theorem~\cite{ET1}, the Goldstone bosons are absorbed
by the gauge bosons to build up the longitudinal degrees of
freedom~\cite{ET}.

Such a theory can be described by an effective Lagrangian, expanded in
the dimensions of the field operators, or equivalently the energy
in momentum space~\cite{EffT}\footnote{For a recent theoretical
summary see Ref.\cite{Han} which includes also triple $W$ production
in the $e^+e^-$ annihilation channels~\cite{WWW}, supplementing the
present analysis.}.  This systematic expansion gives rise to a
parameter-free prediction of the $WW$ scattering amplitudes to leading
order; the leading-order predictions therefore reflect the basic
dynamical mechanism which breaks the electroweak symmetries.
Higher orders in the expansion are determined by the detailed
structure of the underlying new strong-interaction theory.  The
effective Lagrangian can, in unitary gauge, be written as
\begin{equation}\label{L}
  \LL = \LL_g 
  + \LL_0 + \LL_4 + \LL_5 + \ldots
\end{equation}
$\LL_g$ describes, in standard notation~\cite{WW1}, the kinetic terms of
the gauge fields:
\begin{equation}\label{Lg}
  \LL_g = -\frac18 \tr{W_{\mu\nu}^2} - \frac14 B_{\mu\nu}^2
\end{equation}
$\LL_0$, the lowest-order term in the chiral expansion, corresponds to
the mass terms:
\begin{equation}\label{L0}
  \LL_0 = M_W^2 W^+_\mu W^-_\mu + \frac12 M_Z^2 Z_\mu Z_\mu
\end{equation}
The two terms, $\LL_g+\LL_0$, generate the parameter-free $WW$
scattering amplitudes to leading order in the energy region where the
$WW$ interactions become strong.  The dimension-4 operators $\LL_4$
and $\LL_5$ are new quadrilinear contact interactions of the $W^\pm$
and $Z$ bosons:
\begin{align}\label{L45}
  \LL_4 &= \alpha_4\left[      
	\frac{g^4}{2}\left[(W^+_\mu W^-_\mu)^2 
	+ (W^+_\mu W^+_\mu)(W^-_\nu W^-_\nu)\right]
        + \frac{g^4}{c_w^2}(W^+_\mu Z_\mu)(W^-_\nu Z_\nu) 
	+ \frac{g^4}{4c_w^4}(Z_\mu Z_\mu)^2
        \right]
	\nonumber\\
  \LL_5 &= \alpha_5\left[
        {g^4}(W^+_\mu W^-_\mu)^2 
	+ \frac{g^4}{c_w^2}(W^+_\mu W^-_\mu)(Z_\nu Z_\nu)
        + \frac{g^4}{4c_w^4}(Z_\mu Z_\mu)^2
        \right]
\end{align}
with $c_w^2=1-\sin^2\theta_w$ and $g^2=e^2/\sin^2\theta_w$.
$\alpha_4$ and $\alpha_5$ are the parameters of the next-to-leading
order terms in the expansion.  These contact terms introduce
all possible quartic couplings compatible with the custodial $SU(2)_c$
symmetry.  The amplitudes for the $WW$ scattering processes may be
expressed in terms of a master amplitude $A$ which is a function of the
Mandelstam variables $s$, $t$ and $u$:
\begin{align}
  A(W^+W^-\to ZZ) &= A(s,t,u) \label{aZZ}\\
  A(W^+W^-\to W^+W^-) &= A(s,t,u) + A(t,s,u) \label{aWW}\\
  A(W^-W^-\to W^-W^-) &= A(t,s,u) + A(u,t,s) \label{asWW}
\end{align}
The dominating strong-interaction part of the master amplitude is
given by the expansion
\begin{equation}
  A(s,t,u) = \frac{s}{v^2} + \alpha_4\frac{4(t^2+u^2)}{v^4}
        + \alpha_5\frac{8s^2}{v^4}
\end{equation}
with $v^2=1/(\sqrt2 G_F) = (246\ \GeV)^2$.  The leading-order term
$s/v^2$ of the expansion is parameter free.

It is generally expected that $e^\pm e^-$ linear colliders~\cite{LC} will in
a first step be realized for a total c.m.\ energy up to about
$1~\TeV$, see Refs.\cite{Sitges}.  Moreover, a high
integrated luminosity of $\int\LL = 1~\ab^{-1}$ may be reached within
two years of operation with TESLA.  Since due to the
complicated mixture of signal and background mechanisms, simple
scaling laws are not trustworthy \textit{a priori}, we have updated
the $WW$ scattering analysis of Ref.\cite{WW1} for a total c.m.\
$e^\pm e^-$ energy of $\sqrt{s}=1~\TeV$ and integrated luminosities
of $\int\LL_{e^+e^-}=1~\ab^{-1}$ for $e^+e^-$ collisions, and
$\int\LL_{e^-e^-}=100~\fb^{-1}$ for $e^-e^-$ collisions.  Electron and
positron polarizations are assumed to be $100\%$ and $50\%$, respectively.

\begin{figure}
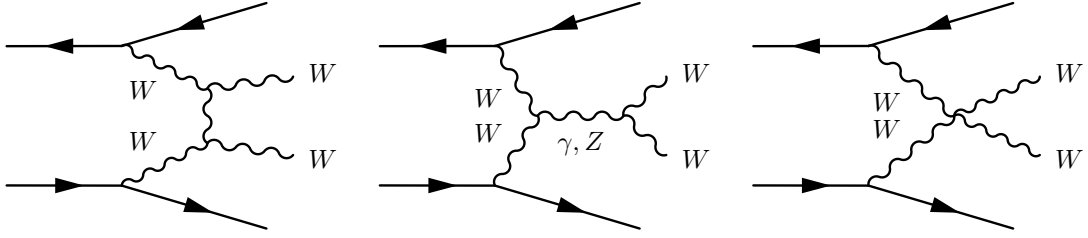

\begin{center}
\unitlength1mm
\begin{picture}(40,30)
\put(0,0){\includegraphics{wwgraphs.1}}
\input{wwgraphs.t1}
\end{picture}
\qquad
\begin{picture}(40,30)
\put(0,0){\includegraphics{wwgraphs.2}}
\input{wwgraphs.t2}
\end{picture}
\qquad
\begin{picture}(40,30)
\put(0,0){\includegraphics{wwgraphs.3}}
\input{wwgraphs.t3}
\end{picture}
\end{center}
\caption{Diagrams contributing to the strong $WW$ scattering signal.}
\label{fig-signal}
\end{figure}

\begin{figure}
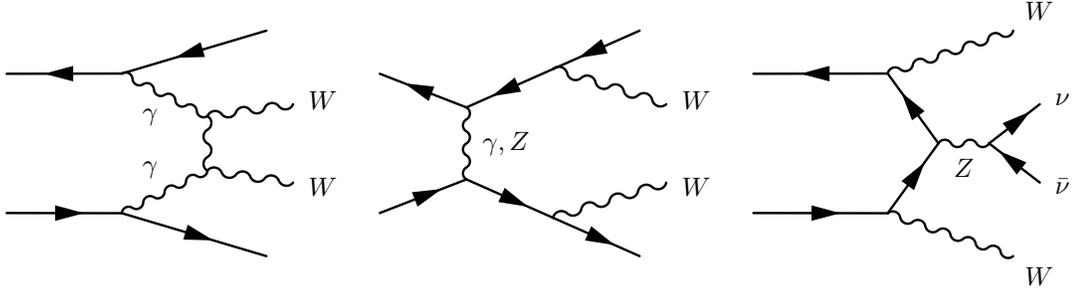

\begin{center}
\unitlength1mm
\vspace*{\baselineskip}
\begin{picture}(40,30)
\put(0,0){\includegraphics{wwgraphs.4}}
\input{wwgraphs.t4}
\end{picture}
\qquad
\begin{picture}(40,30)
\put(0,0){\includegraphics{wwgraphs.5}}
\input{wwgraphs.t5}
\end{picture}
\qquad
\begin{picture}(40,30)
\put(0,0){\includegraphics{wwgraphs.6}}
\input{wwgraphs.t6}
\end{picture}
\end{center}
\caption{Typical diagrams contributing to the background.}
\label{fig-bg}
\end{figure}

\paragraph{} Using the Lagrangian of Eqs.~(\ref{L}--\ref{L45}), the
cross sections have been determined for the processes
\begin{align}
  e^+e^- &\to \bar\nu\nu W^+W^- \quad\text{and}\quad \bar\nu\nu ZZ\\
  e^-e^- &\to \nu\nu W^-W^-
\end{align}
by calculating the amplitudes analytically and performing the phase
space integrations numerically.  The analysis includes the signal
diagrams Fig.\ref{fig-signal} as well as all relevant background
diagrams (a few important examples are depicted in
Fig.\ref{fig-bg}). 


The strategy for isolating the signal from the background has been
described in Ref.\cite{WW1} in detail.  For the present analysis we have used
the following cuts on the final-state particles:
\begin{center}
\begin{minipage}{10cm}
\vspace{.5\baselineskip}
\begin{itemize}
  \item[$\mathcal{C}$:]
  $M(\nu\bar\nu)>150\;\text{GeV}$
  \item[]
  $|\cos\theta(W/Z)| < 0.8$ and 
  $p_\perp(W/Z) > 100\;\text{GeV}$
  \item[]
  $p_\perp(WW) > 40\;\text{GeV}$ resp. $p_\perp(ZZ)>30\;\text{GeV}$
  \item[]
  $400\;\text{GeV} < M(WW/ZZ) < 800\;\text{GeV}$
\end{itemize}
\vspace{.5\baselineskip}
\end{minipage}
\end{center}
The efficiency for the detection of vector bosons and the probability
of $W/Z$ misidentification are determined by the decay branching
ratios and by the detector resolution for invariant jet pair masses.
Taking into account both leptonic and hadronic decays, we adopt the
numbers from Ref.\cite{WW1} which amount to an overall detection
efficiency of $33\%$ for both $WW$ and $ZZ$ pairs in the final state.

\begin{figure}[p]
\begin{center}
\includegraphics{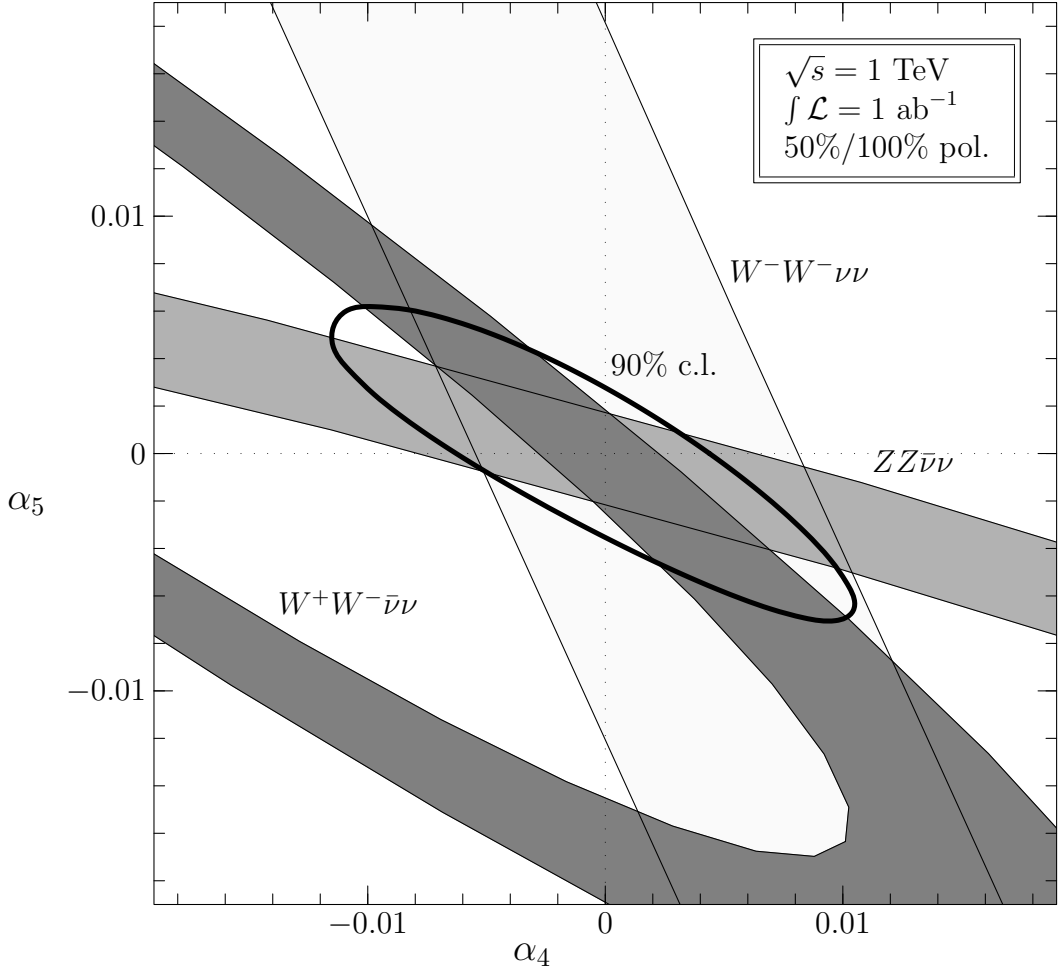}
\end{center}
\vspace{\baselineskip}
\caption{Exclusion contours for the hypothesis $\alpha_{4,5}=0$, assuming
$\protect\sqrt{s}=1\ \TeV$ and an integrated $e^+e^-$ luminosity of
$\int\LL=1\ \ab^{-1}$ ($50\%/100\%$ polarization).  The $90\%$
exclusion line has been obtained by combining the $W^+W^-$ and $ZZ$
channels (dark gray).  The contour for the $W^-W^-$ channel (light gray)
corresponds to an integrated $e^-e^-$ luminosity of $\int\LL=100\
\fb^{-1}$ ($100\%$ polarization).}
\label{contour}
\end{figure}

\paragraph{}
The results of this analysis are summarized in Fig.\ref{contour}.
Exclusion contours at the $1\sigma$ level are shown for the parameters
$[\alpha_4,\alpha_5]$ as derived from the three processes introduced
above.  The highest sensitivity is predicted for the $W^+W^-$ and $ZZ$
channels; the additional $W^-W^-$ channel, however, is useful for
resolving the two-fold ambiguity and singling out the unique solution.
For an energy of $1~\TeV$ and luminosities as specified above, the
dynamical parameters $\alpha_4$ and $\alpha_5$ can be measured to an
accuracy
\begin{align}
  \alpha_4 &\lesssim 0.010 \\
  \alpha_5 &\lesssim 0.007
\end{align}
When compared with the results of Ref.\cite{WW1} for higher energy but
reduced luminosity, $\alpha_{4,5}\leq 0.002$, the bounds follow
\emph{roughly} the scaling law $\alpha_{4,5}\propto s^{-1}\times
(\int\LL)^{-1/2}$ which may be used for qualitative inter- and
extrapolations.  As a threshold effect, the sensitivity improves
dramatically with rising energy.

Assuming the same scaling law in luminosity also for LHC analyses~\cite{LHC}
one finds bounds on $\alpha_4$ and $\alpha_5$ which are about a
factor~$2.5$ and $3$ less stringent after two years of high-luminosity
running for a total equivalent of $\int\LL=200~\fb^{-1}$, and provided
the systematic errors can be kept under control at this level.
Nevertheless, the correlation between the parameters in individual
channels is different so that independent information can be obtained
from experiments at lepton and hadron colliders.

The sensitivity bounds on $\alpha_{4,5}$ can be rephrased in bounds on
the errors with which the lowest-order part of the master amplitude
\begin{equation}
  A(s,t,u)_{\textrm{LO}} = {s}/{v^2}
\end{equation}
can be determined experimentally\footnote{These experimental analyses
will only be carried out in the future for a physical scenario in
which light Higgs bosons have experimentally been proven not to exist.
The comparison of $WW$ scattering amplitudes between theories without
and with light Higgs bosons is therefore a \textit{res vacua} in this
specific context.}.  Taking proper account of the angular dependence
of the coefficients coming with $\alpha_4$ and $\alpha_5$, the
accuracy on the master amplitude is given by
\begin{equation}
  \langle\delta A/A\rangle \lesssim 0.15
\end{equation}
for an average $WW$ invariant mass of $\sim 600~\GeV$, corresponding
to a total $e^+e^-$ energy of $1~\TeV$, and an integrated luminosity of
$\int\LL = 1~\ab^{-1}$. 

Thus high-luminosity $e^+e^-$ colliders allow us to test the basic
mechanism for electroweak symmetry breaking even in the absence of a
light Higgs boson quite stringently at a collider energy of $1~\TeV$.

\subsubsection*{Acknowledgement}
We thank F.~Richard for encouraging the analysis presented in this
addendum.

\vspace{2cm}



\end{document}